\documentclass[12pt]{iopart}
\renewcommand{\tableofcontents}{}

\usepackage{graphicx}
\usepackage{cite}
\usepackage{xcolor}

\usepackage{subcaption}
\usepackage{acronym} 
\acrodef{ML}{Machine Learning}
\acrodef{PF}{Particle Flow}
\acrodef{PFA}{Particle Flow Algorithm}
\acrodef{PID}{Particle Identification}
\acrodef{HCAL}{Hadronic Calorimeter}
\acrodef{DHCAL}{Digital Hadronic Calorimeter}
\acrodef{SDHCAL}{Semi-Digital Hadronic Calorimeter}
\acrodef{AHCAL}{Analogue Hadronic Calorimeter}
\acrodef{RPWELL}{Resistive Plate WELL}
\acrodef{RPC}{Resistive Plate Chambers}
\acrodef{MM}{MicroMegas}
\acrodef{NN}{Neural Network}
\acrodef{GNN}{Graph neural network}
\acrodef{GAT}{Graph Attention Transformer}
\acrodef{MLP}{multilayer perceptron}
\acrodef{LHC}{Large Hadron Collider}
\acrodef{MIP}{Minimum Ionizing Particle}
\acrodef{RPC}{Resistive Plate Chamber}
\acrodef{HEP}{High Energy Physics}
\acrodef{TP}{Truth Positive}
\acrodef{FP}{False Positive}
\acrodef{TN}{Truth Negative}

\newcommand\cmt[1]{}

\usepackage{gensymb}

\begin{document}

\title{Exploring DHCAL design  and performance with Graph Neural Networks}

\author{M. Borysova$^{1}$, D. Zavazieva$^{1,2}$, N. Kakati$^{1}$, E. Gross$^{1}$ and S. Bressler$^{1}$}

\address{$^{1}$Weizmann Institute of Science, Rehovot, Israel}
\address{$^{2}$Ben-Gurion University of the Negev, Be'er Sheva, Israel}
\vspace{10pt}

\date{October 2024}

\begin{abstract}
In the context of a gas-sampling \ac{DHCAL}, we explore the potential of using \acp{GNN} for hadron energy reconstruction and \ac{PID} in future collider experiments. For \ac{PID}, we achieved classification efficiencies exceeding 50\% for neutrons and pions, with notably higher efficiencies for kaons and protons. Protons exhibited the highest efficiency of 77\%, followed by neutral kaons. The energy resolution for these hadrons is studied in the energy range of 1 -- 50 GeV, with a further investigation into the resolution as a function of the incoming particle's angle and readout granularity, focusing on charged pions. Compared to traditional analysis methods, our results indicate that improved performance can be achieved even with coarser detector granularity, potentially making future \ac{DHCAL} systems more cost-effective.

\end{abstract}

\tableofcontents

\acresetall 

\section{Introduction}
\label{sec:introduction}

\begin{figure}[ht!]\centering
    \begin{subfigure}{0.44\linewidth}\centering
        \includegraphics[width=\linewidth]{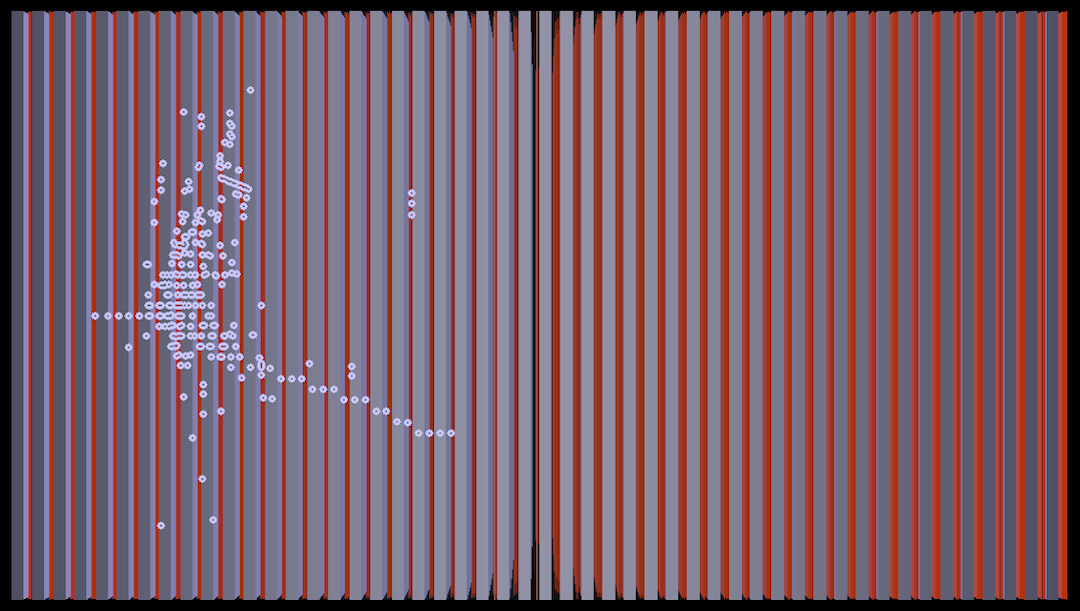}
        \caption{Pion}\label{pion_shower}
    \end{subfigure}
    \begin{subfigure}{0.44\linewidth}\centering
        \includegraphics[width=\linewidth]{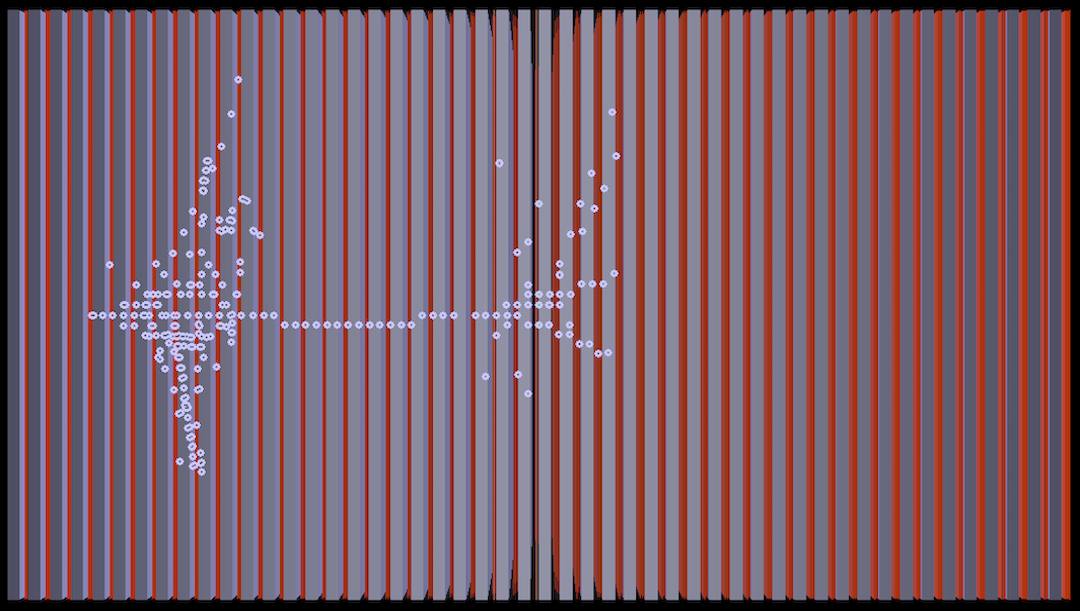}
        \caption{Proton}\label{}
    \end{subfigure}
    \begin{subfigure}{0.44\linewidth}\centering
        \includegraphics[width=\linewidth]{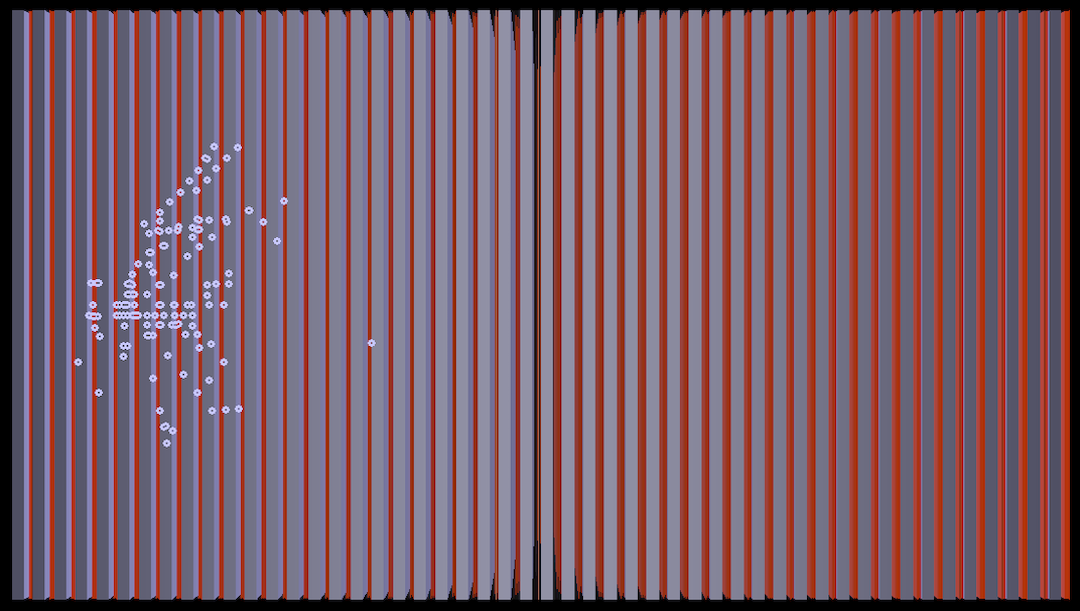}
        \caption{Neutron}\label{}
    \end{subfigure}
    \begin{subfigure}{0.44\linewidth}\centering
        \includegraphics[width=\linewidth]{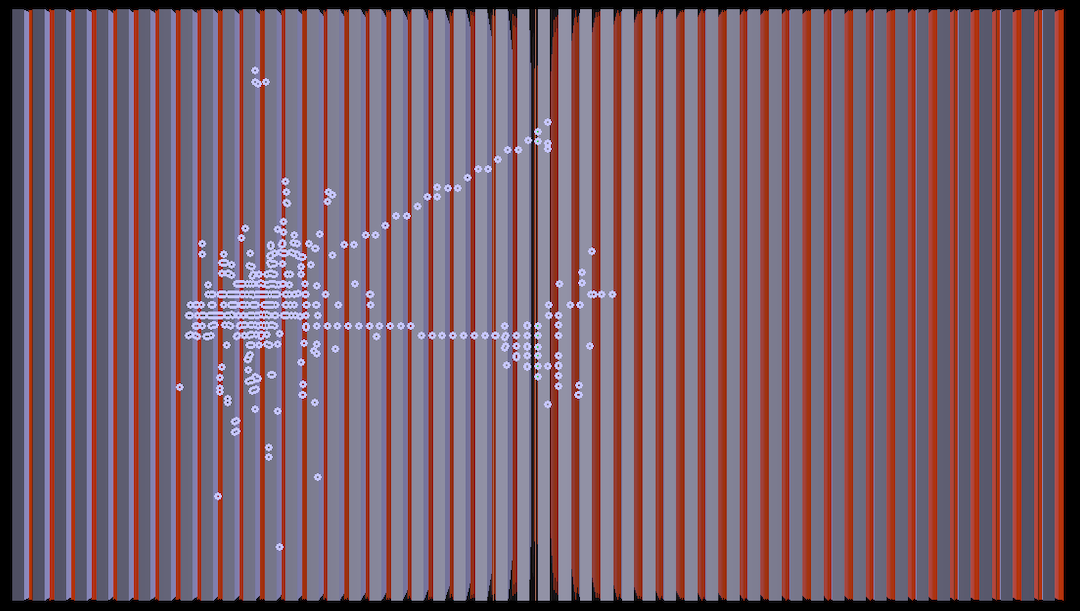}
        \caption{Kaon}\label{}
    \end{subfigure}
   
  \caption{Event displays of simulated $\pi^{-},p, n, K_{L}^{0}$ hadron showers. See Section \ref{sec:G4} for simulation details.}
  \label{fig:particle_showers}
\end{figure}

\ac{ML} has made a profound impact on physics research, particularly in particle physics, where it has been successfully applied to a wide range of tasks, including data collection, physics object reconstruction, and \ac{PID}. In particle physics, accurate jet measurement is essential for understanding the fundamental properties of particles and their interactions. Therefore, the \ac{HCAL} in future collider facilities requires exceptional hadron energy resolution, with the goal of achieving $\frac{\Delta E}{E} \leq 55 \%$~\cite{Thomson:2009rp}. 
Despite advancements, precision measurement of hadronic showers remains a significant challenge. When hadrons interact with a calorimeter, a diverse array of nuclear processes can occur as the energy is absorbed. As demonstrated in Figure \ref{fig:particle_showers}, this leads to substantial event-by-event fluctuations in the types and multiplicity of secondary particles, the spatial distribution of their energy deposits, and the fraction of invisible energy lost to nuclear binding energy. These factors are all dependent on the energy and type of the incident particle.

One of the most significant challenges arises from the differing responses of most \acp{HCAL} to hadrons and electrons, i.e., e/h ratio is not equal to unity~\cite{Wigmans}. This discrepancy complicates the task of directly measuring the incoming hadron energy based solely on the detected signals. This problem is enhanced due to the large event-by-event fluctuations of hadronic showers which are attributed, among others, to the energy transfer from hadronic to electromagnetic component via $\pi_{0}$ production. 

The energy reconstruction of these showers can be further complicated by energy leakage due to insufficient coverage of the detector or energy losses due to missing channels.
Given the multidimensional challenges associated with reconstructing the energy of hadronic showers, traditional algorithms such as the weighted summation method~\cite{CMS:2022jvd} are often complex, require arbitrary parameterization choices, and do not fully address the aforementioned issues.

\ac{PF} has emerged as a solution to address these challenges and is now the most widely used technique for reconstructing individual particles within jets in high-energy physics experiments~\cite{Thomson:2009rp}. Its goal is to measure each final-state particle within the optimal sub-detector. This requires high granularity in the calorimeters to accurately associate energy deposits with individual particles, separate nearby particle showers, and match the showers of charged particles with tracks in the tracking system.

Several high-granularity \ac{HCAL} systems, which differ in their readout schemes, are considered: \ac{AHCAL}, which measures both the position and the deposited energy; \ac{SDHCAL}, incorporating typically $1\times 1~cm^2$ pads with 2-bit precision to measure the deposited energy, potentially providing information on hit multiplicity; and \ac{DHCAL}, which features a simpler binary readout (hit/no-hit) for $1\times 1~cm^2$ pads. The latter approach simplifies the complexity and cost of the calorimeter's readout system, making smaller cell sizes practical for a hadron calorimeter.

Traditional \ac{PFA} methods convert the number of measured hits in the \ac{DHCAL} to the energy of the incoming particle~\cite{Shaked-Renous:2022kxo, CALICE:2019rct}. More advanced algorithms also account for the hit density in different layers. \ac{ML} models offer a robust approach for developing customized shower separation algorithms based on event-specific information. Currently, experiments at the CERN \ac{LHC} and future circular colliders utilize parameterized \ac{PFA}~\cite{DiBello-2021} with sophisticated energy clustering algorithms~\cite{DiBello-2021, Pata-2024, Simkina:2023cmx} and with inclusion of timing information~\cite{CMSHGCAL:2023rsx}.

\acp{GNN} have emerged as an architecture of choice in recent particle reconstruction models, demonstrating superior performance in shower separation compared to traditional convolutional neural networks for \ac{PF}~\cite{DiBello:2022iwf}. This success is primarily due to their ability to learn the shower shape within a given detector geometry.

\ac{ML}-based reconstruction algorithms, assuming \acp{AHCAL} with a highly granular scintillator medium, have been developed by the CALICE~\cite{CALICE:2024imr}, CMS~\cite{CMSHGCAL:2024esz}, and CEPC~\cite{Song:2023ceh} collaborations. CALICE has also explored such algorithms for \ac{SDHCAL}, using Glass \acp{RPC} as the sensitive medium~\cite{SDHCAL-2020}.

In this work, we explore the potential of \acp{GNN} to improve the design and performance of a sampling \ac{HCAL} with digital (1 bit) readout, i.e., \ac{DHCAL}. Building on the approach of~\cite{Shaked-Renous:2022kxo}, we use a GEANT4~\cite{GEANT4:2002zbu} model to simulate the response of a fully equipped \ac{RPWELL}-based \ac{DHCAL} to pions, protons, kaons, and neutrons at a variety of energies, angles, readout granularity and performances. In the following we discuss  the application of \acp{GNN} to energy reconstruction and \ac{PID}. The methods employed in this work are detailed in Section \ref{sec:methods}, followed by the results in Section \ref{sec:results}. We discuss our findings in Section \ref{sec:discussion}.
\section{Methods}
\label{sec:methods}

\subsection{GEANT4 simulation}
\label{sec:G4}

\begin{figure}[ht]
\centering
\begin{minipage}{.5\textwidth}
  \begin{subfigure}{\linewidth}
    \centering
    \includegraphics[width=.8\linewidth]{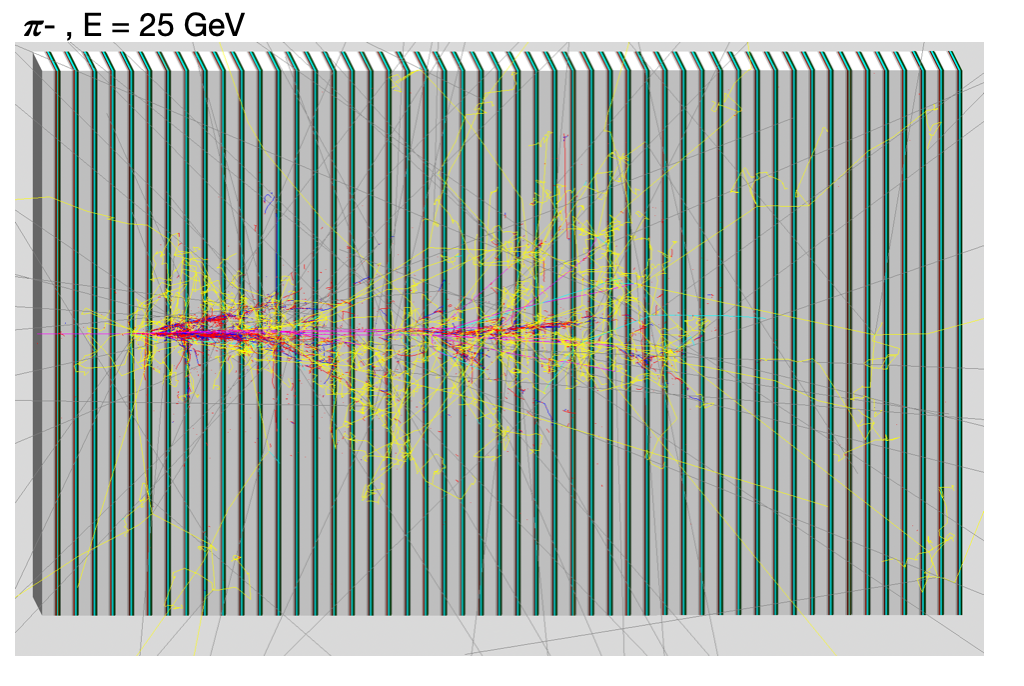}
    \caption{}
    \label{fig:calo50}
  \end{subfigure}
  \begin{subfigure}{\linewidth}
    \centering
    \includegraphics[width=.8\linewidth]{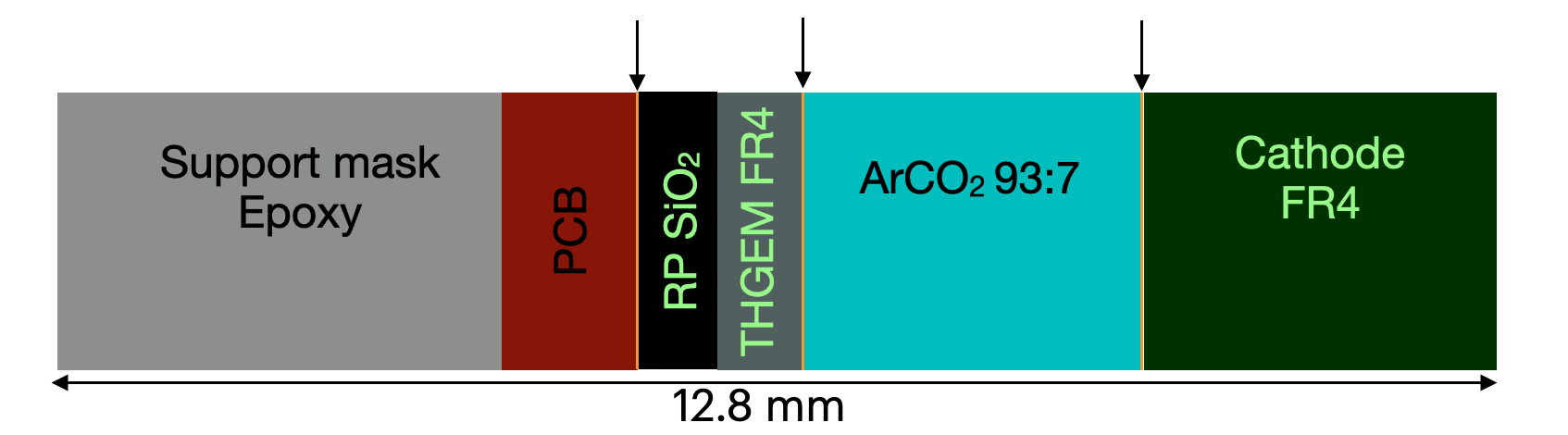}
    \caption{}
    \label{fig:module}
  \end{subfigure}
\end{minipage}%
\begin{minipage}{.5\textwidth}
  \begin{subfigure}{\linewidth}
    \centering
    \includegraphics[width=.8\linewidth]{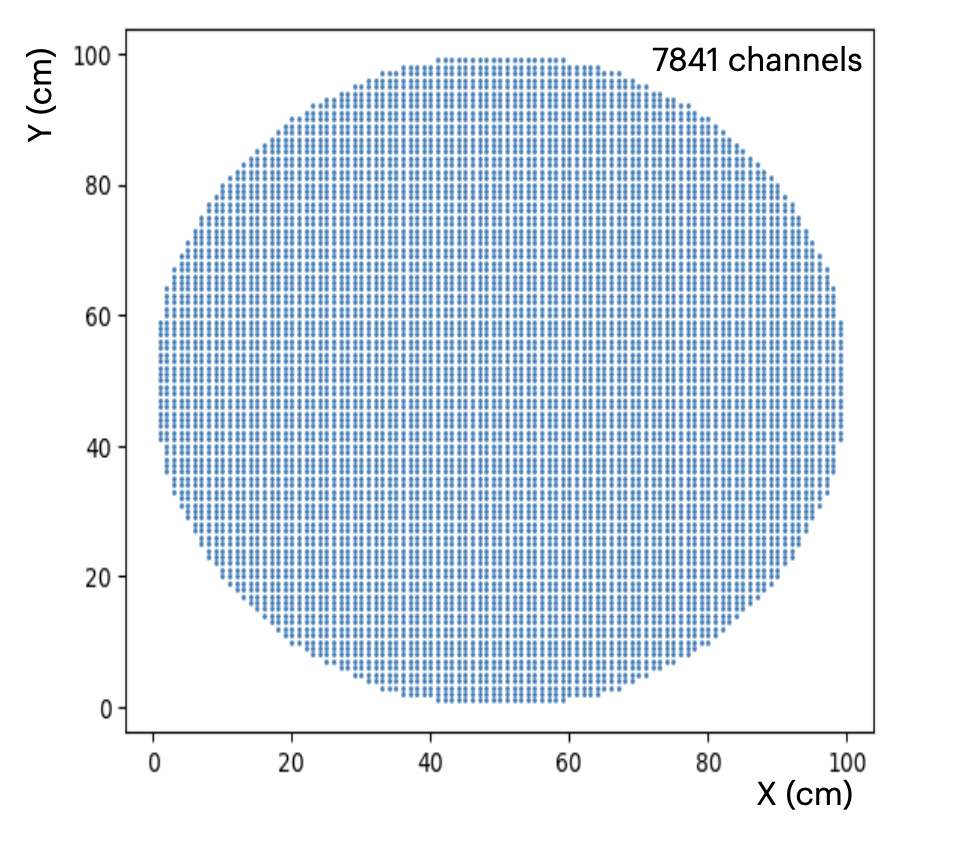}
    \caption{}
    \label{fig:readout}
  \end{subfigure}
\end{minipage}%

\caption{The 50 layers \protect{\ac{DHCAL}} GEANT4 module detailed in~\cite{Shaked-Renous:2022kxo}. (a) A shower of a 25 GeV pion, excluding photons. (b) A \protect{\ac{RPWELL}}-based sampling element; the small arrows point from left to right on the readout anode, \protect{\ac{RPWELL}} to electrode, and copper cathode. (c) Readout anode with $1\times 1~cm^2$ pads. }
\label{fig:DHCAL_G4}
\end{figure}

GEANT4\footnote{version 10.06.p01 \cite{GEANT4:2002zbu}, with QGSP-BERT-EMZ physics list} was used to model a \ac{DHCAL} module and generate datasets for training and testing the performance of the \acp{NN}. The model was extensively validated with test beam data in \cite{Shaked-Renous:2022kxo}. It consists of 50 layers of \ac{RPWELL}-based sampling elements and 2 cm-thick steel absorbers (Figures~\ref{fig:calo50}, \ref{fig:module}), corresponding to a total depth of approximately $5\lambda_{\pi}$. This depth ensures a 99.3\% probability that a pion will initiate a shower within the module, minimizing energy leakage. To mitigate further the effects of longitudinal leakage, we applied a pre-selection criterion, requiring the shower to initiate within the first 10 layers of the calorimeter. Depending on the simulated hadron, this condition was met by 64-67\% of all simulated showers.  

Sampling elements with readout pads arranged in a circular pattern, featuring varying pad sizes ranging from $1\times 1~cm^2$ to $4\times 4~cm^2$, were considered to explore the sensitivity of the performance on the readout granularity. A complete anode design using $1\times 1~cm^2$ pads is shown in Figure~\ref{fig:readout}. The energy deposits were digitized into electronic signals, emulating various pad multiplicities and hit detection efficiency values. Fired neighboring pads were grouped into clusters, with the cluster's position determined as the average of the positions of all individual pads within the cluster.

As detailed in Section \ref{sec:results}, depending on the specific task, different datasets were generated. These include different mixtures of hadrons, their energies, and impinging angles. For each particle type, the data set, after pre-selection, contains approximately 600k events that were used for the training of the \acp{NN}, 100k events for validation and another set of 110k events, not seen by the network, for performance testing. For the angle studies, larger data sets were generated: 6M (training) and 1.5M (validating/testing) events for testing performance in the angular range of 0-40$\degree$ and 1.8M, 450k for narrower ranges of angles.

\subsection{Neural Networks} 
\label{sec:NN}


In \ac{HEP}, data is often heterogeneous and sparse, with numerous inter-dependencies, making graph-based algorithms a natural choice \cite{Shlomi_2021}. The calorimeter data under investigation is similarly sparse, with shower profiles encoded in the relationships between hits, motivating our decision to employ a graph-based representation.

\cmt{
\subsubsection{Hadronic showers as graphs}
\label{sec:showersAsNN} 
}

While the calorimeter data can be interpreted as a 3D image, traditional image-based methods are not well-suited to address the inherent sparsity and would not generalize well to detectors with different geometries.

\begin{figure}[ht]
\centering
\begin{minipage}{.5\textwidth}
  \begin{subfigure}{\linewidth}
    \centering
    \includegraphics[width=.8\linewidth]{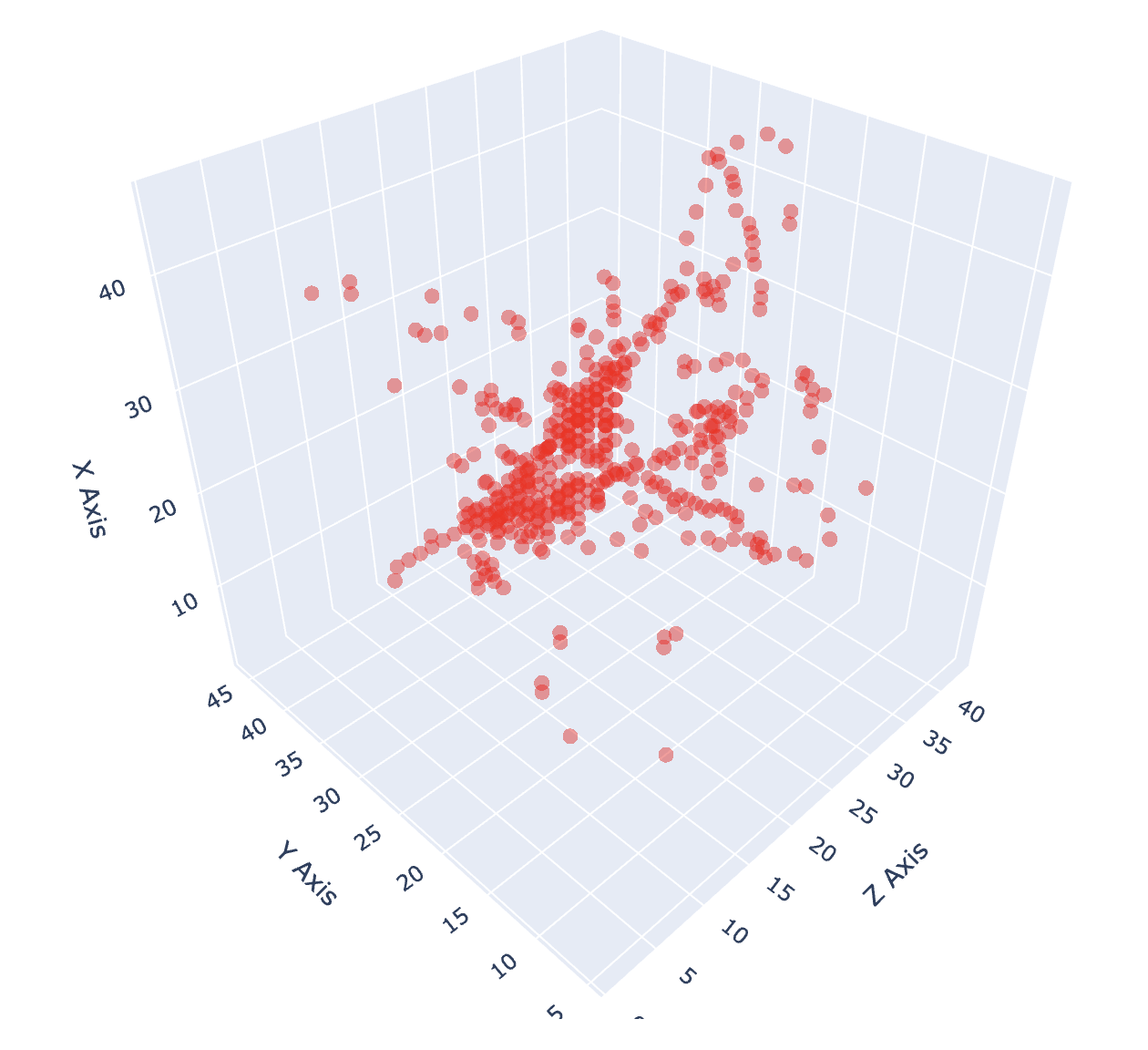}
    \caption{}
    \label{fig:point_cloud}
  \end{subfigure}
\end{minipage}%
\begin{minipage}{.5\textwidth}
  \begin{subfigure}{\linewidth}
    \centering
    \includegraphics[width=.8\linewidth]{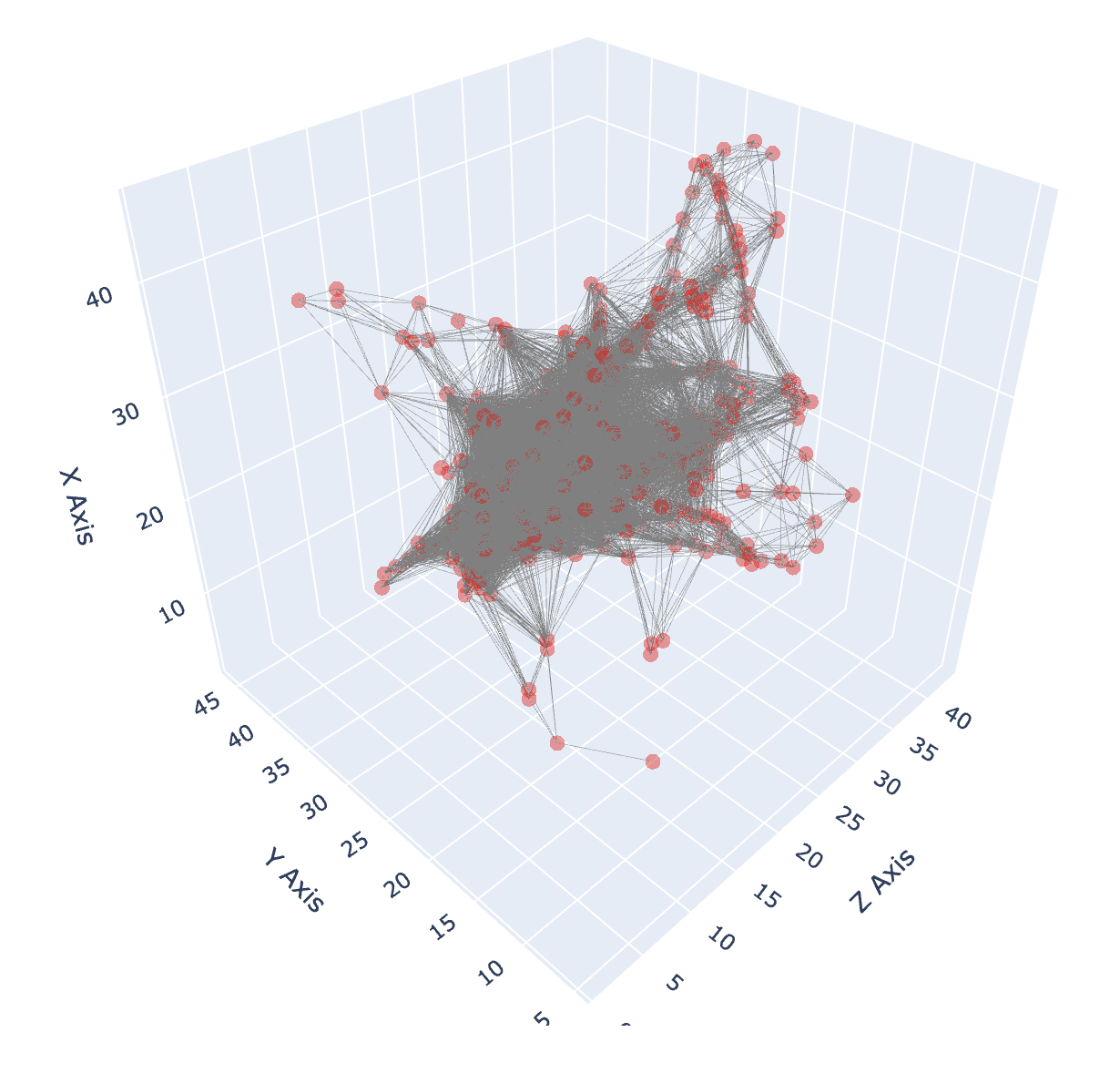}
    \caption{}
    \label{fig:graph_radius}
  \end{subfigure}
\end{minipage}
\caption{Representations of pion showers as a point cloud (a) and as a graph connecting nodes in a rectangular box spanning over a spatial volume of 10 pads in the X and Y directions and 10 layers in the Z direction (b). }
\label{fig:graph_pion_shower}
\end{figure}

Relying on the graph-based approach, we represent each calorimeter cluster as a node in the graph. A collection of disconnected nodes forms a point cloud, whereas nodes connected by edges define a graph. We investigated two types of connectivity: k-nearest neighbor and radius-based methods. In this case, the latter method defines a rectangular box which connects nodes within a spatial volume of 10 clusters in all three directions. It was found to provide the most effective connectivity. The node features are characterized by the $x$, $y$, and $z$ coordinates of the cluster. Figure~\ref{fig:graph_pion_shower} illustrates two resulting graphs for a single charged pion shower: the point cloud configuration (Figure~\ref{fig:point_cloud}) and the graph with connected nodes (Figure~\ref{fig:graph_radius}). In addition to the individual nodes, the total number of nodes in the shower was also provided as an input to the \ac{NN}.

\begin{figure}[ht]
\centering
\includegraphics[width=0.95\linewidth]{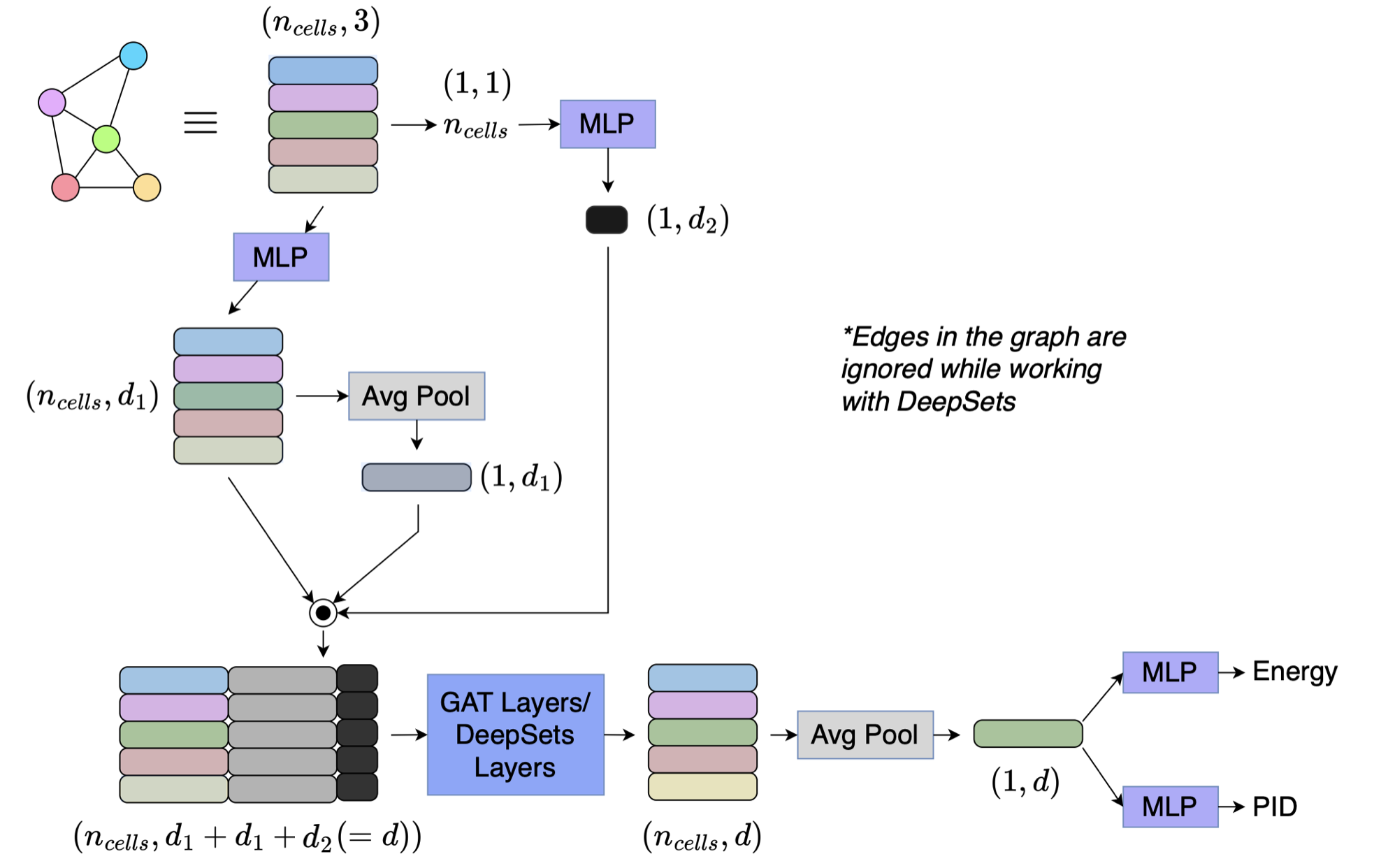} \\
\caption{The NN architectures. Both the DeepSets and the GAT-based approaches share a similar overall structure, differing only in the use of GAT layers versus DeepSets layers. The network takes $n_{cells}$ activated calorimeter cells as input (colored circles), each with 3 initial spatial features (colored ovals) encoded by a \ac{MLP}. An average pooling layer (grey oval) aggregates these features. The total cell count (black oval) is concatenated with the aggregated features, resulting in a vector of dimension $d$. This vector is processed by GAT or DeepSets layers, averaged (sage green oval), and passed through a MLP for energy or PID prediction.}
\label{fig:nn_architectures}
\end{figure}

\cmt{
\subsubsection{DeepSets and GATs}
\label{sec:NNArchitecture}
}
In this study, we explored two \ac{NN} architectures: one based on DeepSets ~\cite{zaheer2018deepsets} and the other utilizing \acp{GAT}~\cite{yuan2024graphattentiontransformernetwork},~\cite{vaswani2023attentionneed}. Figure \ref{fig:nn_architectures} illustrates both architectures. In each case, the input nodes, representing calorimeter clusters, are passed through a \ac{MLP} that encodes the features into a higher-dimensional space. 
These cell representations are then refined through multiple iterations of either DeepSets or \acp{GAT} layers.

The input to the DeepSets architecture is a point cloud. In each iteration, the nodes are aggregated using average pooling, which is then employed to update the individual cluster features. Conversely, the input to the \acp{GAT} are nodes with edges (graphs). Special attention is given to the edges, allowing the \acp{GAT} model to effectively exploit the underlying structure of the shower, potentially improving performance. As illustrated in Fig.~\ref{fig:nn_architectures}, both architectures share a similar overall structure. The GAT model utilizes four layers, while the DeepSets architecture employs ten DeepSets layers, with these configurations chosen based on empirical performance during experimentation.

After several updates using either \acp{GAT} or DeepSets, the refined features of all clusters are averaged to create a global representation. This global representation encapsulates the overall energy deposition of the event. A final \ac{MLP} then processes this global feature vector to generate the final predictions, which are either energy estimates or classification logits (raw scores used for classification tasks).

Both architectures were implemented in PyTorch~\cite{pytorch2017Smith}, with the AdamW~\cite{Loshchilov2017DecoupledWD} optimizer utilized for training. Although full hyper-parameter optimization can potentially enhance model performance, we opted for a practical approach, exploring a range of learning rates and batch sizes. A learning rate of $10^{-4}$ is found to provide a good balance between convergence speed and stability. Similarly, a batch size of 64 was chosen as a compromise between computational efficiency and model performance. These settings were sufficient to achieve promising results and demonstrate the effectiveness of our approach.

\subsection{\ac{PID} and energy reconstruction}
\label{PIDErec}

\subsubsection{The \ac{NN} prediction}
\label{sec:targets}

We trained the \acp{NN} for three different targets: predicting particle energy, identifying particle type, and jointly predicting both with equal weights. The \ac{NN} trained to predict particle energy outputs a single numerical value corresponding to the predicted energy. The \ac{NN} trained for \ac{PID} produces a multi-class tensor with n=4 values, where n represents the number of potential incident particle candidates. These values are normalized using the Softmax function~\cite{softmax} and interpreted as probabilities for each candidate. The particle type is then determined by selecting the candidate with the highest probability. Finally, the \ac{NN} trained to predict both particle energy and type provides the two outputs.

\subsubsection{Performance quantification}
\label{sec:quantification}

The energy reconstruction is evaluated by fitting the distribution of $\frac{\sigma_E}{E}$ as a function of the energy of the incoming particle to the commonly used functional form $\frac{S}{\sqrt{E}} \oplus C$, where S and C are the stochastic and constant terms, respectively.

To evaluate the \ac{PID} performance, we used a commonly adopted definition of efficiency in classification tasks. This metric quantifies how accurately the model identifies the target class while avoiding misclassifications. Let \ac{TP} denote the number of events correctly classified as the target class, \ac{FP} the number of events incorrectly classified as the target class, and \ac{TN} the number of events correctly classified as not belonging to the target class. The efficiency ($\varepsilon$) and fake rate ($f$) are defined as follows:

\[ \varepsilon = \frac{TP}{TP + FP} \]

\[ f = \frac{FP}{FP + TN} \]

The efficiency measures the proportion of correctly identified target events relative to all instances classified as the target. A high fake rate indicates that the model frequently misclassifies non-target events as belonging to the target class. Results are often presented in a confusion matrix in which the rows represent the "True" generated particle and the columns the "Predicted" particle. In this matrix the diagonal corresponds to the efficiency value and the off-diagonal to the fake rates.  

\section{Results}
\label{sec:results}

Each of the two \ac{NN} architectures was trained to predict one of three targets
(only energy, only \ac{PID}, both energy and \ac{PID}). The datasets were simulated using the \ac{RPWELL}-based \ac{DHCAL} module described in Section \ref{sec:G4}. Four types of hadrons -- neutrons, negative pions, protons, and neutral kaons ($K^{0}_{L}$) -- were simulated with initial energies uniformly distributed between 1 and 60 GeV. 
The composition of the training dataset vary to better address the question at hand.

\subsection{Particle Identification}
\label{sec:resPID}
The study was carried out assuming 98\% \ac{MIP} detection efficiency (based  on \ac{MIP} detection efficiency measured with sampling elements of different technologies~\cite{Shaked-Renous:2022kxo}) and 1.0 average pad-multiplicity for each sampling element.  Single neutrons, pions, protons, and kaons were simulated to traverse the center of the \ac{DHCAL} module perpendicularly to the XY plane (0-degree incidence angle). Four independent datasets, corresponding to each particle type, were combined into a single training dataset comprising 2.4 million events. Additionally, validation and test datasets containing 400k and 440k showers, respectively, were generated with the same composition.

\begin{figure}[ht]
\centering
\begin{minipage}{.33\textwidth}
  \begin{subfigure}{\linewidth}
    \centering
    \includegraphics[width=.9\linewidth]{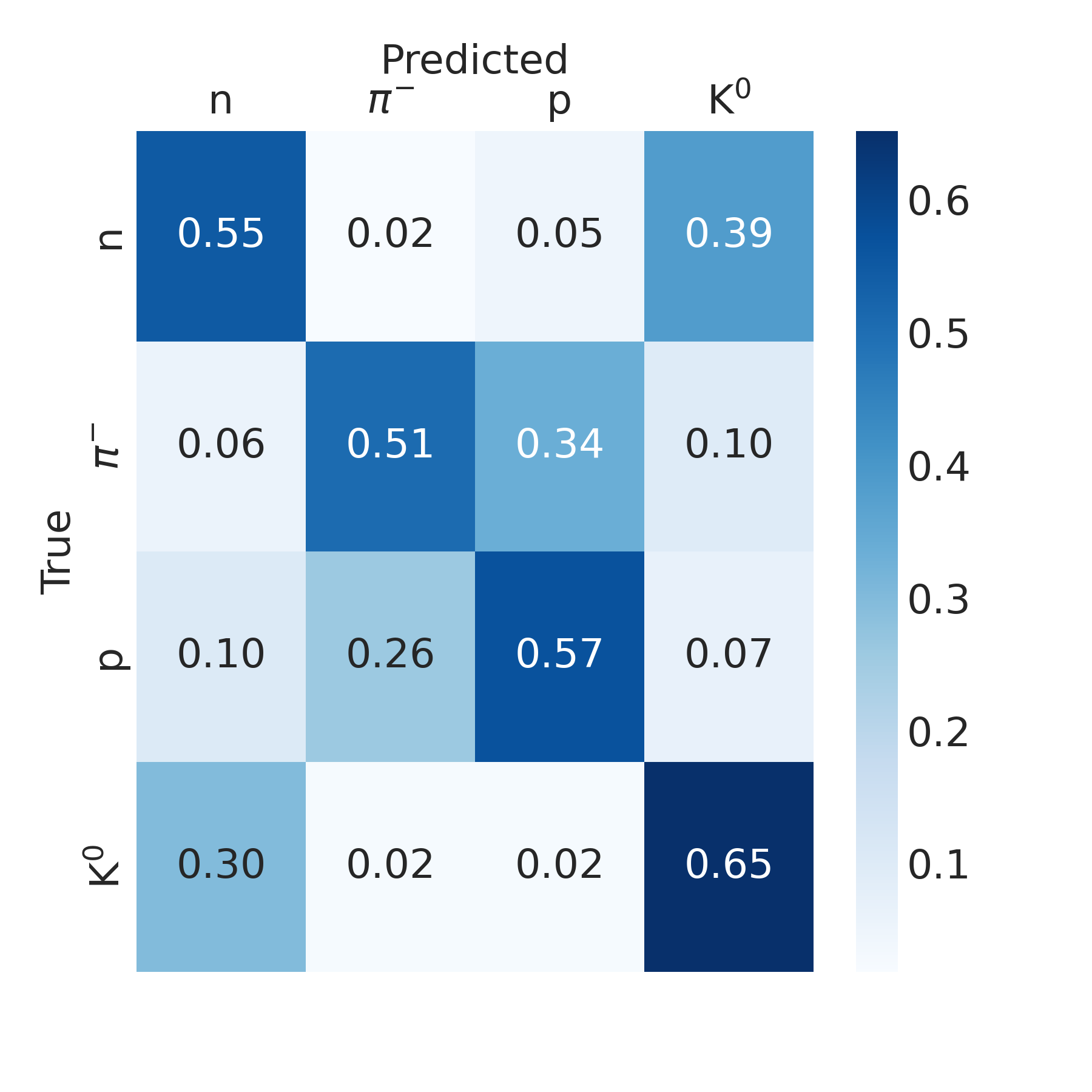}
    \caption{}
    \label{fig:id_gat}
  \end{subfigure}
\end{minipage}%
\begin{minipage}{.33\textwidth}
  \begin{subfigure}{\linewidth}
    \centering
    \includegraphics[width=.9\linewidth]{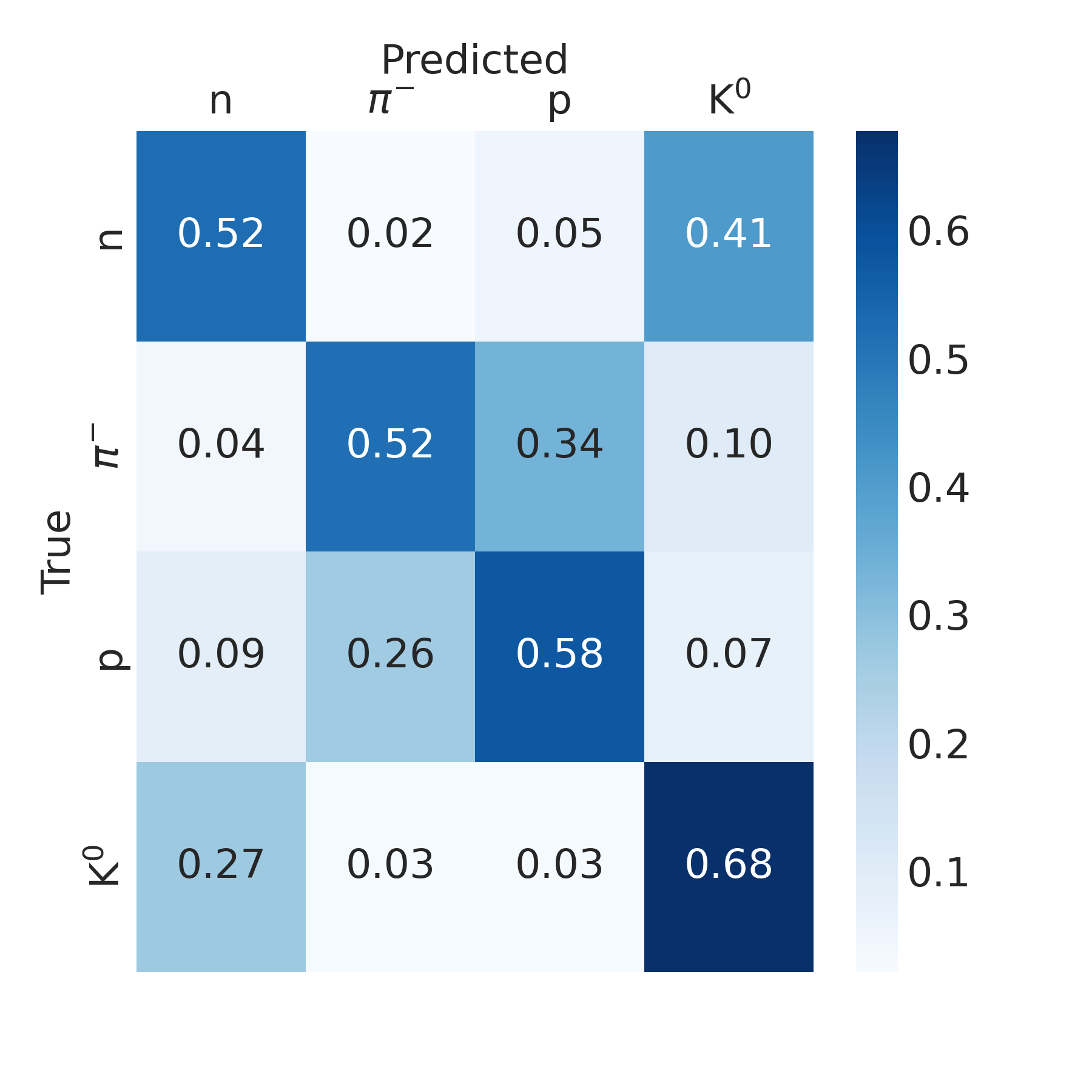}
    \caption{}
    \label{fig:idres_gat}
  \end{subfigure}
\end{minipage}%
\begin{minipage}{.33\textwidth}
  \begin{subfigure}{\linewidth}
    \centering
    \includegraphics[width=.9\linewidth]{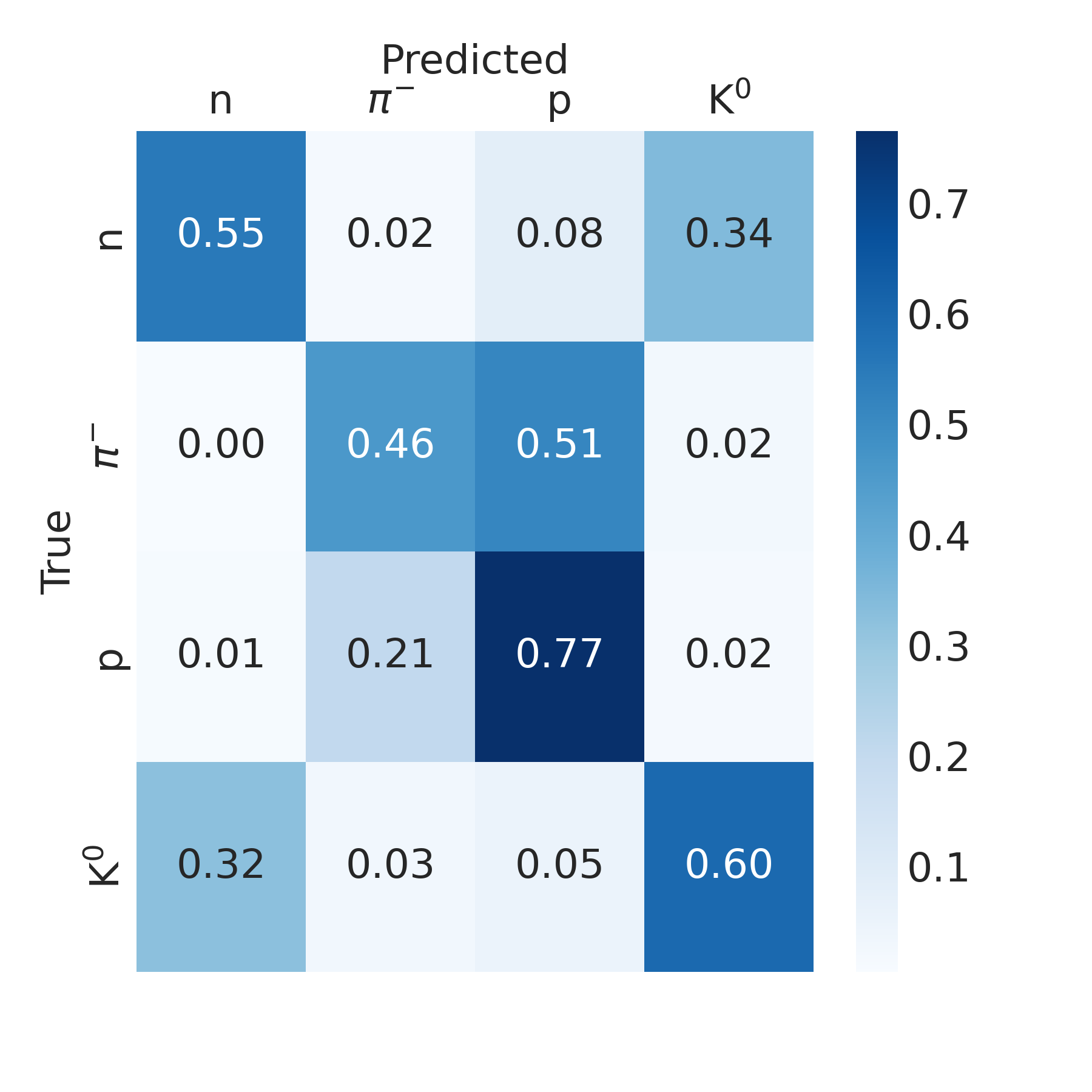}
    \caption{}
    \label{fig:idres_deepset}
  \end{subfigure}
\end{minipage}
\caption{\ac{PID} confusion matrix (see Section \protect\ref{sec:quantification}) predicted with a \ac{GAT} trained to predict only the \ac{PID} (a) and with \ac{GAT} (b) and DeepSets (c) trained to predict both \ac{PID} and the energy.}
\label{fig:idres}
\end{figure}

Figure~\ref{fig:idres} depicts the \ac{PID} efficiency, $\varepsilon$, and fake rate, $f$, obtained with a \ac{GAT} trained to predict only the \ac{PID} (\ref{fig:id_gat}) and with \ac{GAT} (\ref{fig:idres_gat}) and DeepSets (\ref{fig:idres_deepset}) trained to predict both \ac{PID} and the energy at equal weights. An attempt to train DeepSets to predict only the \ac{PID} failed.

As can be seen in Figure~\ref{fig:idres}, all \ac{GNN} models demonstrate \ac{PID} efficiencies exceeding 50\% for pions and neutrons, and at the level of 60-70\% for protons and kaons. The misidentification occurs particularly between pions and protons and between neutral kaons and neutrons. 

DeepSets, when trained to simultaneously predict PID and energy, achieved the highest proton identification efficiency (77\%), but exhibited reduced performance for kaons and pions, while the \ac{GAT} model yielded the highest kaon identification efficiency (68\%).

According to ~\cite{Akchurin-1998}, the better identification of protons and kaons should be attributed to baryon and strangeness number conservation, respectively, limiting the production of $\pi_{0}$s in their showers, and hence the event-by-event fluctuations.  In contrast, incoming $\pi^{\pm}$s are likely to produce either $\pi^{\pm}$s or $\pi_{0}$s during their interactions. Due to the immediate decay to two photons,  the $\pi_{0}$s contribute to an electromagnetic component in the shower. Its magnitude varies significantly depending on the fraction of $\pi_0$s produced and on the stage at which they are first produced — early or late in the hadronic shower. This impacts both the total measured energy and the shower shape. The poorer \ac{PID} recorded with neutrons is attributed to large event-by-event fluctuations due to large variety of nuclear interaction - spallation, elastic and inelastic scattering, n-induced fission, etc. - initiating the shower. 

A consistent misclassification pattern across all models involves neutral particles, which exhibit diffuse, cloud-like shower topologies (Figure~\ref{fig:particle_showers}(c, d)) compared to charged particles with characteristic \ac{MIP} tracks at the shower's origin.
To investigate the impact of the shower topology on  \ac{PID} performance, a test was conducted using the GAT model on a dataset comprising only neutrons and pions.  Increasing the proportion of trackless pions in the training dataset from 14\% to 29\% resulted in a 4\% improvement in pion PID. 

\begin{figure}[ht]
\centering
\begin{minipage}{.49\textwidth}
  \begin{subfigure}{\linewidth}
    \centering
    \includegraphics[width=.9\linewidth]{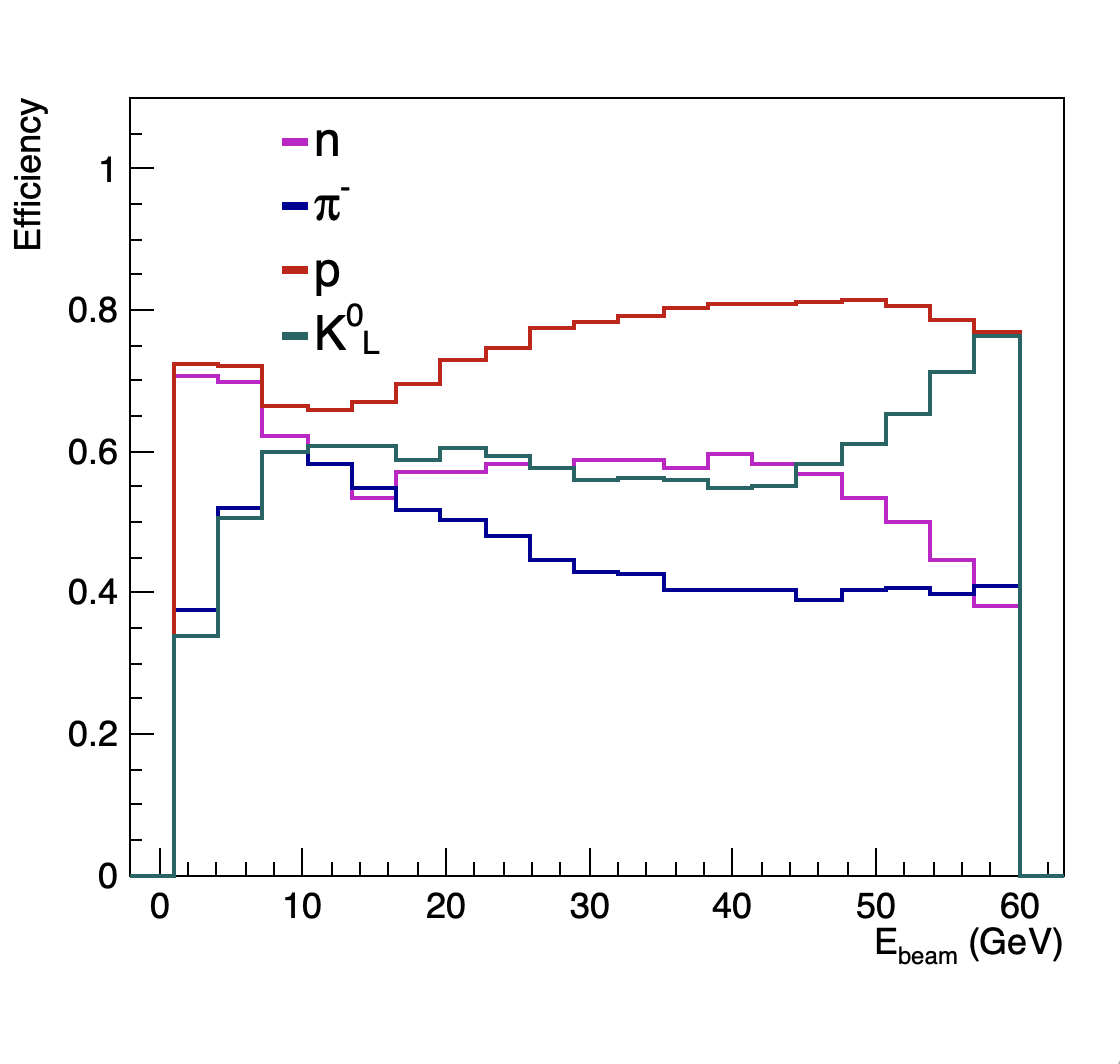}
    \caption{}
    \label{fig:effVsE}
  \end{subfigure}
\end{minipage}%
\begin{minipage}{.49\textwidth}
  \begin{subfigure}{\linewidth}
    \centering
    \includegraphics[width=.9\linewidth]{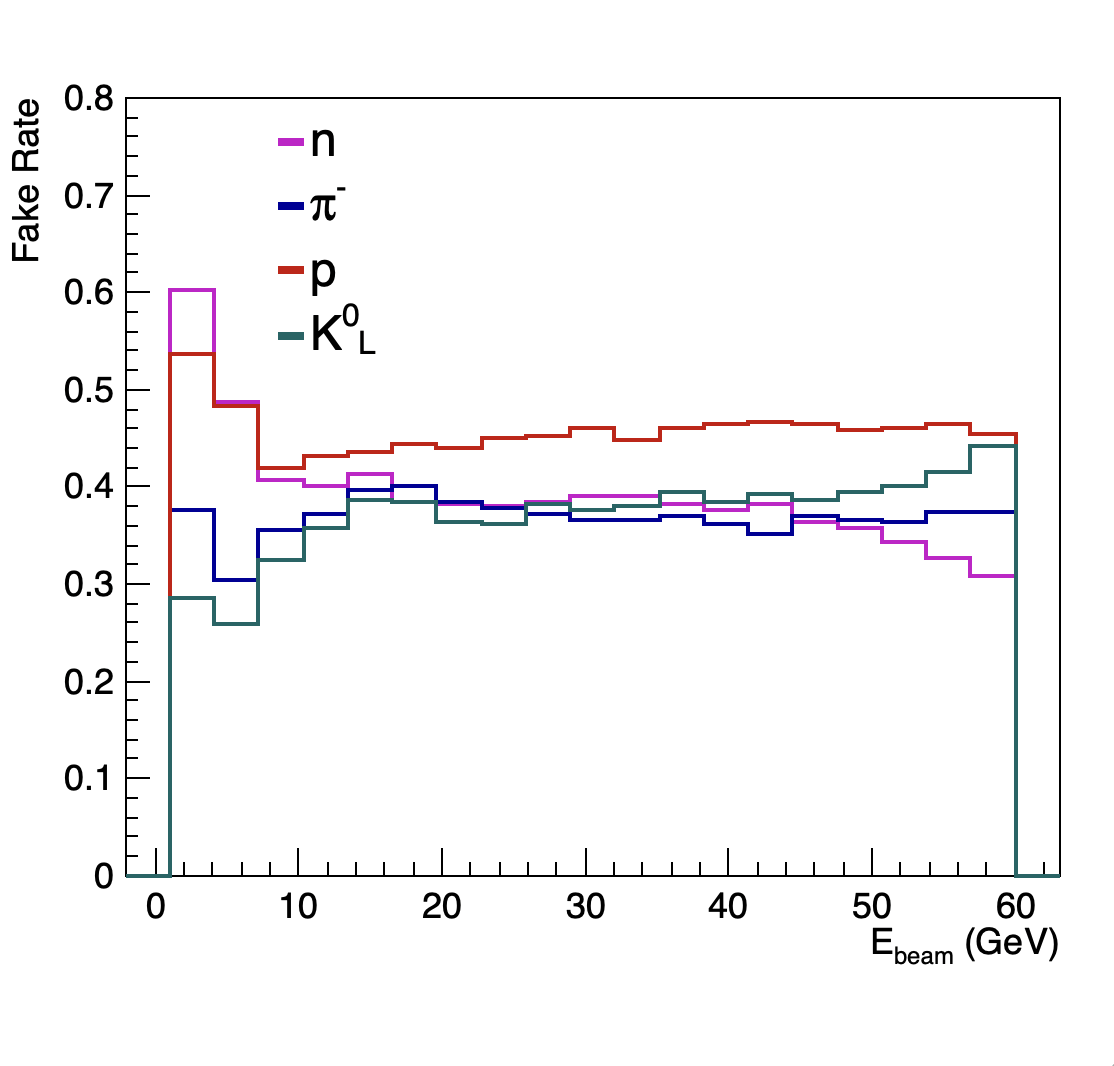}
    \caption{}
    \label{fig:fakeVsE}
  \end{subfigure}
\end{minipage}%
\caption{The \protect{\ac{PID}} efficiency (a) and fake rate (b) as function of the energy of the incoming particle.}
\label{fig:deepset_ResID}
\end{figure}

This suggests that exposing the model to a wider variety of shower patterns, including those without a distinct \ac{MIP} track, improves the \ac{NN} ability to distinguish between pions and neutrons. Additionally, including the total number of clusters in the shower as an input yielded a 2\% improvement in the identification of both neutrons and pions.
For the \ac{GAT} model trained to predict both \ac{PID} and the energy, changing the weight given to each prediction (0.5-0.5, 0.6-0.4, 0.7-0.3) had negligible effect on the identification performance. 
Both models, when trained to simultaneously predict the energy and \ac{PID}, exhibit non-uniform energy dependencies. As illustrated for DeepSets in Figure~\ref{fig:deepset_ResID}(a), the \ac{PID} efficiency for pions and neutrons decreases with increasing energy, while it increases for protons and kaons. Conversely, the fake rates remain relatively consistent, averaging around 40\%, with an elevation at lower energies.
Neutron and proton misclassification rates are particularly elevated below 5 GeV. This is attributed to the similarity in energy deposition patterns between neutrons and protons at lower energies, which poses a challenge for classification algorithms.

\cmt{About $15\%$ of pions faking neutrons are recorded with \ac{GAT}. These are shown to lack a characteristic \ac{MIP} track in the beginning of the shower as in Figure~\ref{fig:particle_showers}(a). The DeepSets architecture trained to predict both the energy and the \ac{PID} demonstrates better performance with overall smaller pion's fake rate ranging from 6\% at small energies and rising up to 12\% at 60 GeV. While neutron's fake rate is higher at energies up to 5 GeV, at which pions mostly do not have the characteristic track in the beginning and falsely classified as neutrons. The efficiencies do not exhibit  strong dependence on the energy of incident particles as depicted in Figure \ref{fig:deepset_ResID}.}

\subsection{Energy reconstruction}
\label{sec:resERec}

For energy reconstruction, negatively charged pions entering the center of the calorimeter perpendicular to its XY plane were studied under various detector response conditions. These include \ac{MIP} detection efficiencies ranging from 90\% to 98\%, and average pad multiplicities of 1.0, 1.1, and 1.6. These values align with those reported in previous measurements of \ac{RPWELL} \cite{Shaked-Renous:2022kxo}, \ac{MM} \cite{Adloff:2013wea}, and \ac{RPC} \cite{CALICE:2019rct} technologies, respectively.

\begin{figure}[!h]
    \centering
    \includegraphics[width=9.cm,clip]{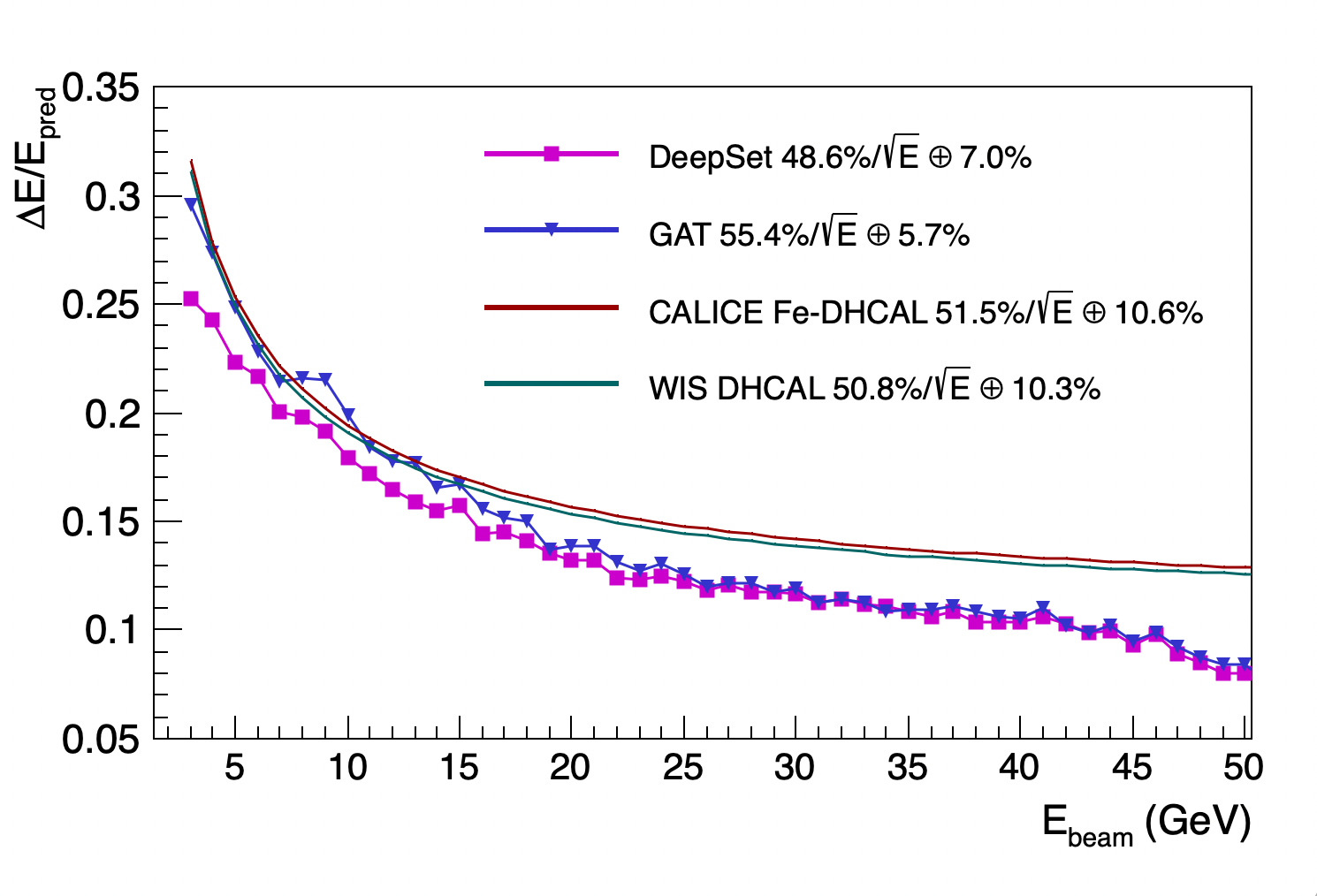}
    \caption{The energy resolution predicted by \protect{\ac{GAT}} and DeepSets in comparison to that obtained with traditional algorithms employed on the same \protect{\ac{RPWELL}}-based \protect{\ac{DHCAL}} and CALICE \protect{\ac{RPC}}-based \protect{\ac{DHCAL}}.}
    \label{fig:res_ds_gat}
\end{figure}

The energy reconstruction performance of DeepSets and \ac{GAT} models, trained solely to predict the energy, is shown in Figure \ref{fig:res_ds_gat}. This is compared with traditional algorithms applied to the \ac{RPWELL}-based \ac{DHCAL} simulation module \cite{Shaked-Renous:2022kxo} and experimental data from the \ac{RPC}-based CALICE \ac{DHCAL} \cite{CALICE:2019rct}. The latter performance was measured up to 32 GeV and extrapolated to 50 GeV in this study. In the low-energy range (up to approximately 15 GeV), DeepSets outperforms all other approaches. Beyond 15 GeV, both DeepSets and \ac{GAT} demonstrate similar performance, surpassing that of traditional algorithms across the energy range.

DeepSets outperforms \ac{GAT} in terms of both the stochastic and constant terms, significantly exceeding the required energy resolution of $55\% / \sqrt E $ for future hadron calorimeters. Furthermore, \ac{GAT} models are considerably more computationally intensive than DeepSets, demanding approximately three times the computational resources and ten times the GPU memory.

\begin{figure}[h]
\centering
\includegraphics[width=9 cm,clip]{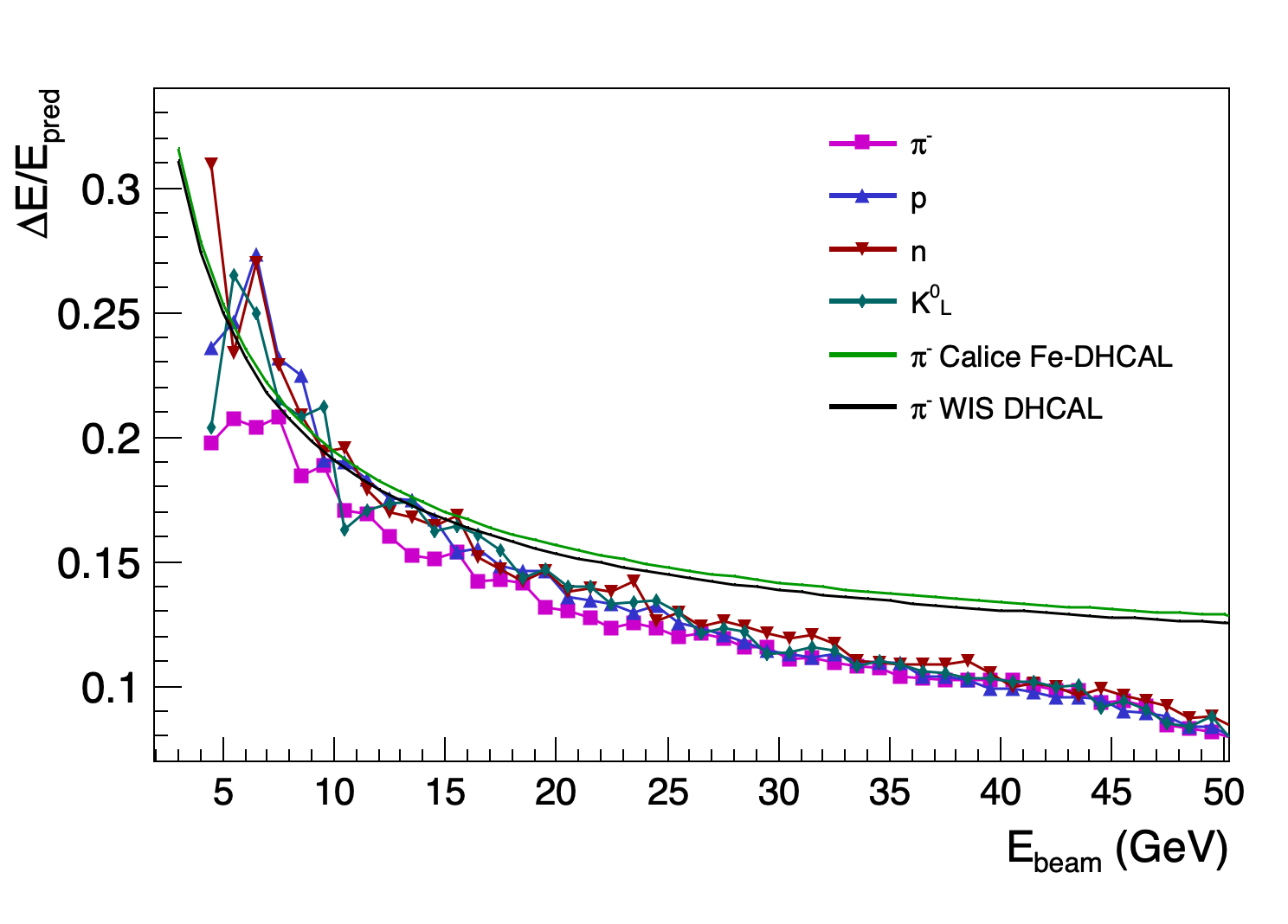}
\caption{The energy resolution of different particles predicted by DeepSets in comparison to that obtained for pions with traditional algorithms employed on the same \protect{\ac{RPWELL}}-based \protect{\ac{DHCAL}}~\cite{Shaked-Renous:2022kxo} and CALICE \protect{\ac{RPC}}-based \protect{\ac{DHCAL}}~\cite{CALICE:2019rct}.}
\label{fig:res_particle}
\end{figure}

To further explore particle-specific performance, four separate DeepSets \acp{NN} were trained, each tailored to predict the energy of a distinct hadron type: pion, neutron, kaon, and proton. The resulting energy resolution for each particle, traversing the calorimeter perpendicularly (0-degree incidence angle), is presented in Figure~\ref{fig:res_particle}. The energy resolution shows minor variation with particle type, with pions exhibiting slightly better resolution than other hadrons up to 25 GeV.
A key characteristic of energy reconstruction, the $E_{pred}/E_{beam}$ ratio versus $E_{beam}$, demonstrates linearity for all hadron types across the presented energy range~\footnote{Excluding very low, $< 5$ GeV, and very high values, $> 50$ GeV, due to known edge effect of the training procedure}.

An \ac{RPWELL}-based \ac{DHCAL} $\mathrm{1\times 1}~m^2$ module with $\mathrm{1\times 1}~cm^2$ readout elements, assuming a \ac{MIP} detection efficiency of 98\% and an average pad multiplicity of 1, was used to investigate the energy resolution of pions as a function of their incident angle. The variation in shower shape and development with the angle of incidence affects the energy deposition patterns within the \ac{DHCAL}, potentially affecting the measured resolution. 

\begin{figure}[!h]
\centering
\includegraphics[width=9cm,clip]{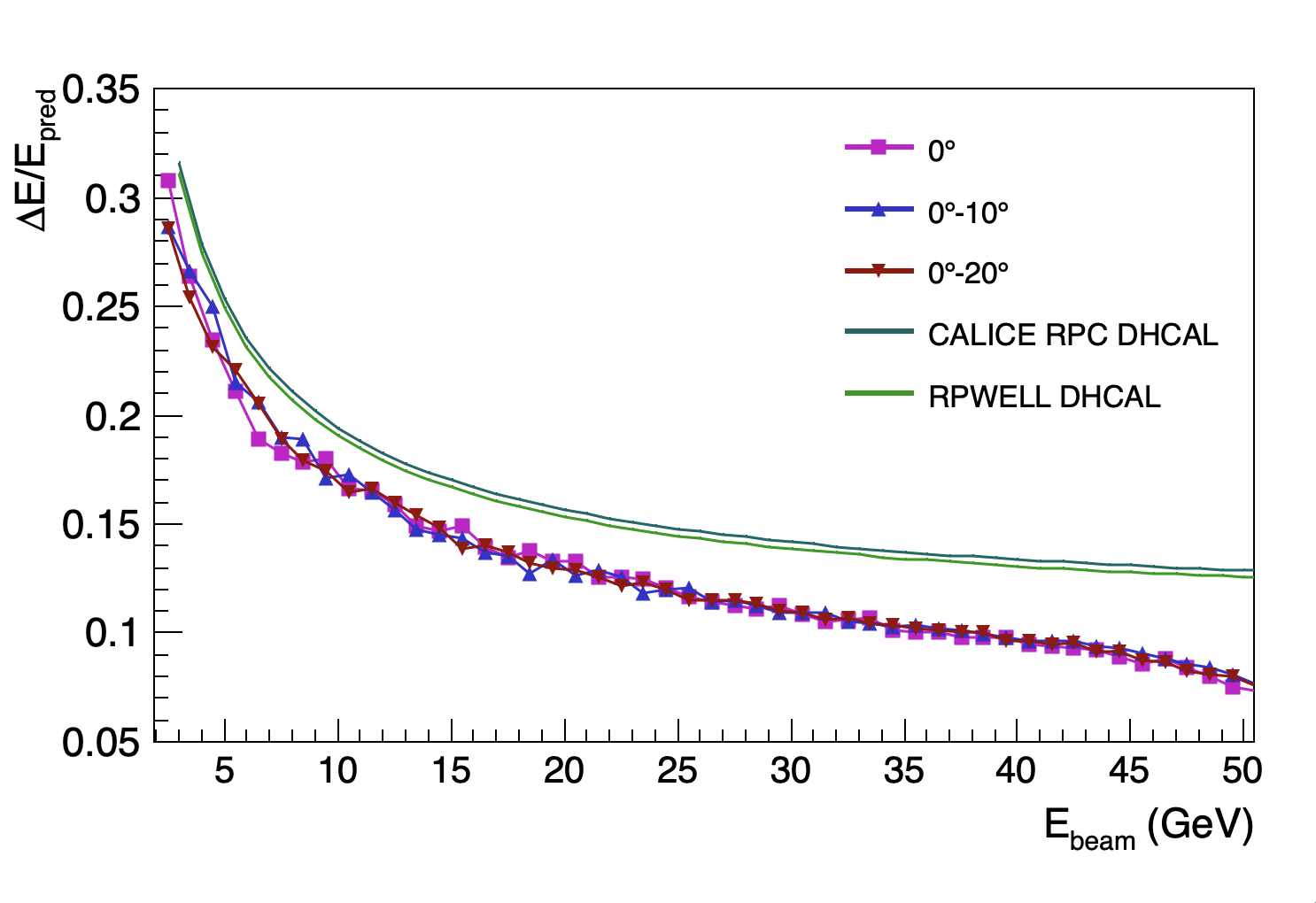}

\caption{The energy resolution of pions predicted by DeepSets trained over different ranges of incoming particles angles.}
\label{fig:res_angles}
\end{figure}

The influence of the incident angle on energy reconstruction was investigated using two training datasets, each containing 1.8 million simulated pion showers. These datasets were generated with pion energies uniformly distributed between 1 and 60 GeV and angles uniformly sampled within $0^{\degree}$–$10^{\degree}$, $0^{\degree}$–$20^{\degree}$, as illustrated in Figure~\ref{fig:res_angles}. 
Comparison of these two datasets revealed no significant degradation in energy resolution up to an incident angle of $20^{\degree}$. Beyond this angle, however, a decrease in resolution is anticipated due to the increased diversity of shower shapes and energy deposition profiles, coupled with enhanced lateral shower leakage from the calorimeter. We plan to expand and potentially refine the training datasets and conduct further studies that incorporate the prediction of the incident angle alongside energy reconstruction.

\subsection{Different detector performance and design}
\label{sec:resDetPerfDes}

We investigated the sensitivity of \acp{GNN} models to various DHCAL detector parameters, including pad size, pad multiplicity, and \ac{MIP} detection efficiency. 
Pion training and testing datasets were generated for various readout granularity and detector performances as shown in Table~\ref{tab:modules}.

\begin{table}[!h]
    \centering
    \begin{tabular}{c|c|c|c}
    \hline
    Module    & $\varepsilon_{MIP}$ & multiplicity & pad size\\
    \hline
    $M_{RPWELL}$  \cite{Shaked-Renous:2022kxo} & 98\%         & 1.1 &     $1 \times 1~\mathrm{cm^2}$ \\                
    $M_2$   & 98\%                & 1.1 &     $2 \times 2~\mathrm{cm^2}$ \\
    $M_3$   & 98\%                & 1.1 &     $3 \times 3~\mathrm{cm^2}$ \\                
    $M_4$   & 98\%                & 1.1 &     $4 \times 4~\mathrm{cm^2}$ \\
    \hline
    $M_5$   & 95\%                & 1.1 &     $1 \times 1~\mathrm{cm^2}$ \\ 
    $M_6$   & 90\%                & 1.1 &     $1 \times 1~\mathrm{cm^2}$ \\ 
    \hline
    $M_{RPC}$ \cite{CALICE:2019rct}  & 96\%    & 1.6 &     $1 \times 1~\mathrm{cm^2}$ \\ 
    \hline

    \end{tabular}
    \caption{List of simulated modules. $M_{RPWELL}$ and $M_{RPC}$ correspond to the \ac{RPWELL}-based \ac{DHCAL} and CALICE-Fe-\ac{DHCAL} performance studied in \cite{Shaked-Renous:2022kxo} and \cite{CALICE:2019rct}, respectively. }
    \label{tab:modules}
\end{table}

\begin{figure}[ht]
\centering
\begin{minipage}{.505\textwidth}
  \begin{subfigure}{\linewidth}
    \centering
    \includegraphics[width=.98\linewidth]{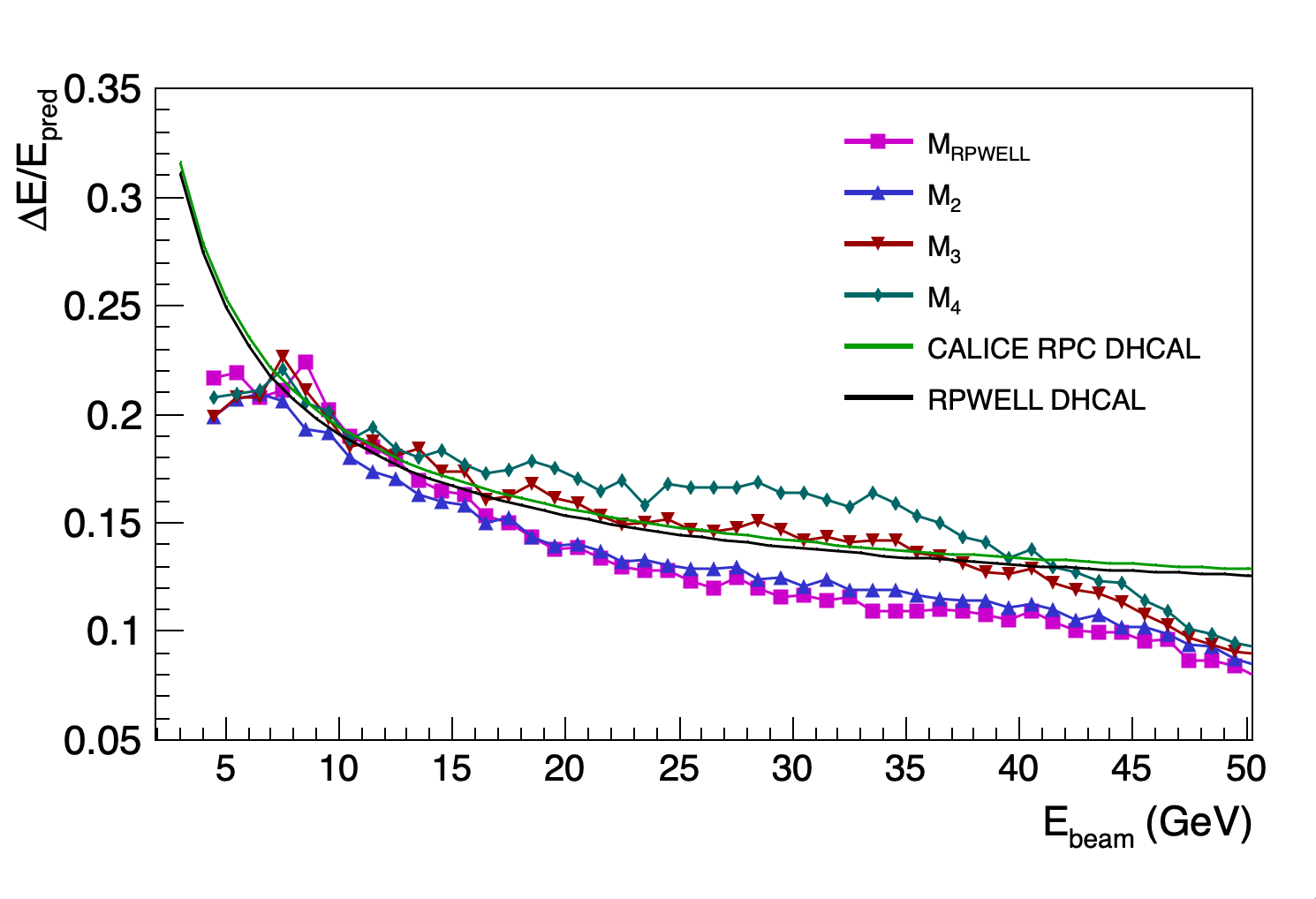}
    \caption{}
    \label{fig:padsize}
  \end{subfigure}
\end{minipage}%
\begin{minipage}{.508\textwidth}
  \begin{subfigure}{\linewidth}
    \centering
    \includegraphics[width=.98\linewidth]{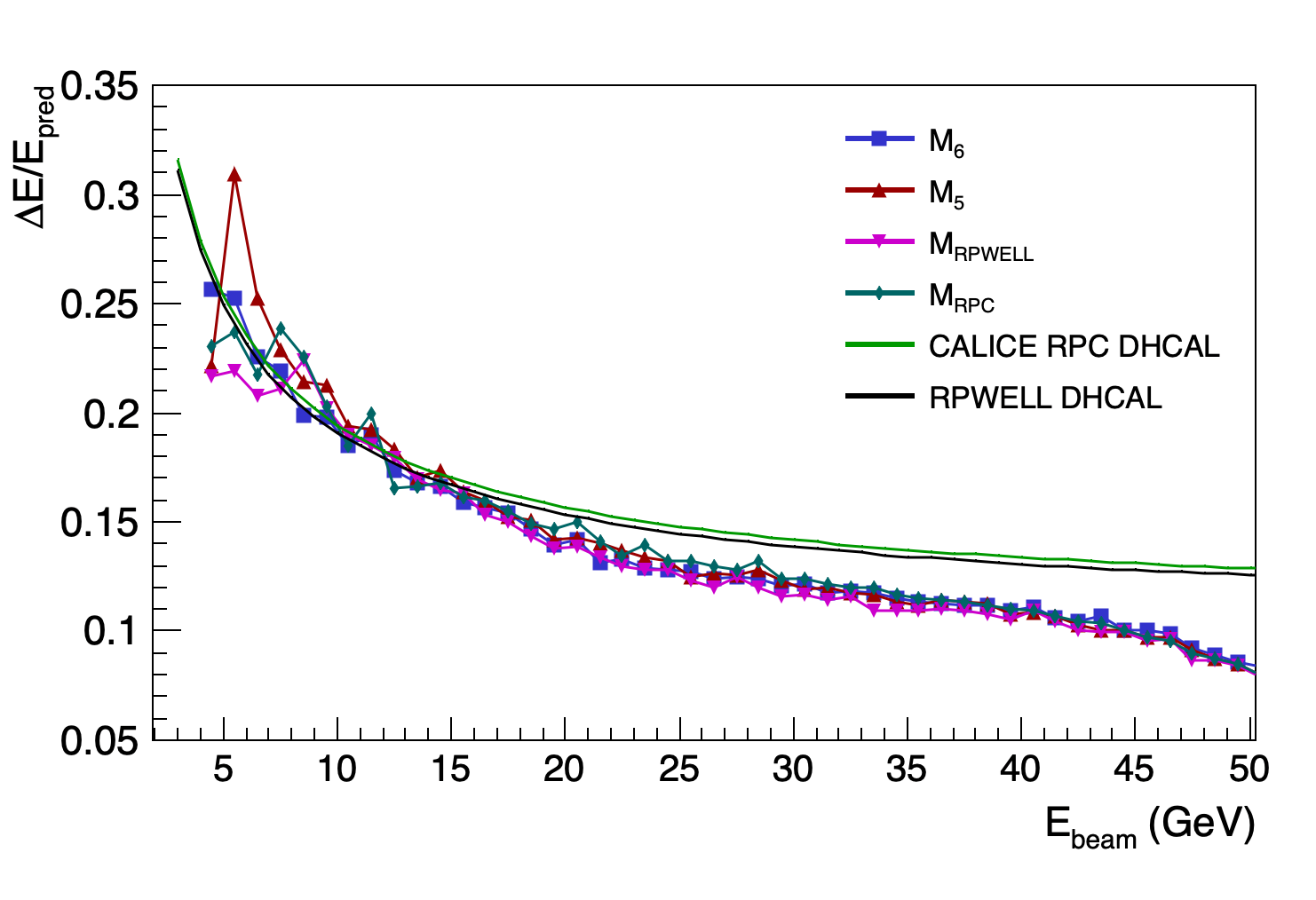}
    \caption{}
    \label{fig:eff}
  \end{subfigure}
\end{minipage}
\caption{(a) The energy resolution of pion predicted by DeepSets for different readout pad sizes. (b) The energy resolution of pions predicted by GAT for varying \protect{\ac{MIP}} detection efficiencies and average pad multiplicities.}
\label{fig:res_pad_eff}
\end{figure}

The baseline design of \ac{DHCAL} module is pad size of $1 \times 1~\mathrm{cm^2}$. 
The dependency of the pion energy resolution for various pad sizes is shown in Figure~\ref{fig:padsize}. 
It demonstrates that enlarging the pad's size by a factor of four ($2 \times 2~cm^2$) and correspondingly reducing the number of channels by four does not degrade the performance significantly. 
 Provided that the two shower separation would not degrade as well, these may offer more cost-effective solution for future experiments. Additionally, the model demonstrated robustness to variations in \ac{MIP} detection efficiency, as illustrated in Figure~\ref{fig:eff}. The detector modules simulated with lower \ac{MIP} efficiencies — the M5 and M6 modules with efficiencies of 90\% and 95\%, respectively — do not significantly impact the resolution. This allows for less stringent restrictions on gain settings, facilitating improved performance without compromising the overall detection capabilities.

\section{Discussion}
\label{sec:discussion}

We investigate potential improvements in the design and performance of future \ac{DHCAL} systems using \ac{GNN} algorithms. These algorithms exploit the representation of hadronic showers as point clouds or connected graphs, offering a more detailed analysis of the shower structure. Multiple DeepSet and \ac{GAT} architectures were trained to predict two objectives — either separately or in combination: the energy and type of the incoming particle. The performance of these \acp{NN} was compared both to each other and to traditional reconstruction algorithms, which primarily rely on counting the number of hits and re-weighting them based on their density. Unlike \ac{GNN}-based approaches, traditional methods do not fully leverage the detailed spatial and structural information of the shower shape.

Training DeepSets and \ac{GAT} to predict the \ac{PID} and energy together, a \ac{PID} efficiencies exceeding 50\%  were achieved for pions and neutrons. DeepSets yielded a peak proton identification efficiency of 77\%, while \ac{GAT} achieved a maximum kaon identification efficiency of 68\%. Across all particle types, the fake rate showed minimal dependence on particle energy. However, proton identification efficiency increased with energy, while pion efficiency decreased, exhibiting energy dependence in those cases.
No model demonstrated a consistently superior performance across all particle types. Future work will focus on refining these models to potentially achieve improved and more uniform \ac{PID} accuracy.

Compared to standard energy reconstruction algorithms, all tested \acp{NN} demonstrated improved pion energy resolution, with the best performance achieved across the entire energy range by the DeepSets model trained exclusively for energy prediction; stochastic term of 48.6\% and constant term of 7\%. The minor discrepancies in energy resolution among different hadrons was found with slightly better energy resolution for pions at small incident energies. 
Enhancing energy resolution at lower energies requires further investigation into the effects of diverse training strategies and network architectures, aiming to improve both the accuracy and robustness of the model.

The relatively poorer performance observed with pions and neutrons, compared to protons and kaons, in both \ac{PID} and energy reconstruction is consistent with phenomenological models that attribute the disparity to fluctuations in the shapes of their particle showers. Addressing this challenge require further research aimed at improving the performance of pions and neutrons, e.g., enhancing the training dataset with low energy neutrons or with trackless pions to better distinguish them from neutrons. 

DeepSets was trained on datasets featuring pion showers with a uniform range of incoming particle angles to evaluate the impact of angle variation on energy resolution. A performance degradation was observed when the \ac{NN} was trained on a broader angular range ($0^{\degree}$–$30^{\degree}$). This degradation is likely due to the increased diversity of shower shapes at each energy, which could be partially mitigated by expanding the size of the training dataset. Further studies to resolve this degradation are ongoing. 

In terms of pion energy resolution, a \ac{DHCAL} module with a pad size of $2 \times 2~\mathrm{cm}^2$ performs as good as a module featuring a pad size of $1 \times 1~\mathrm{cm}^2$. Furthermore, the energy resolution achieved using the DeepSets algorithm on showers recorded with a \ac{DHCAL} module with a $3 \times 3~\mathrm{cm}^2$ pad size was comparable to that obtained using traditional algorithms applied to data from a module with $1 \times 1~\mathrm{cm}^2$ pads. These results suggest that simpler and more cost-effective \ac{DHCAL} designs, with one-fourth or even lower granularity, could suffice for future experiments without compromising performance. Only minor dependency of the pion energy resolution on the \ac{MIP} detection efficiency and average pad multiplicity was found, potentially posing weaker requirements on the performance of the sampling elements in future \ac{DHCAL} systems. Future studies on two-track separation, overall jet-energy resolution, and off-axis showers are crucial for such future decision making.

\section{Acknowledgments}

This study is supported by the Minerva Foundation with funding from the Federal German Ministry for Education and Research, as well as by the Krenter-Perinot Center for High-Energy Particle Physics. Additional support comes from a research grant provided by Shimon and Golde Picker, the Nella and Leon Benoziyo Center for High Energy Physics, and the Sir Charles Clore Prize. Special thanks go to Martin Kushner Schnur for his invaluable support of this research.

\section{References}
\bibliographystyle{unsrt}
\bibliography{references}

\end{document}